\documentclass[sigconf,balance=false]{acmart}
\usepackage{braket}
\usepackage{float}
\usepackage{tikz}
\usepackage{enumitem}
\usetikzlibrary{arrows.meta,positioning,shapes.geometric,calc}

\AtBeginDocument{%
  }

\renewcommand\footnotetextcopyrightpermission[1]{} % removes copyright footnote
\makeatletter
\def\@mkboth#1#2{} % disable mark setting
\let\markboth\@gobbletwo
\let\markright\@gobble
\AtBeginDocument{%
  \pagestyle{plain}
  \thispagestyle{plain}
}
\makeatother

\begin{document}

    \title{QSpy: A Quantum RAT for Circuit Spying and IP Theft}
    
    \author{Amal Raj}
    \email{amal.raj@singaporetech.edu.sg}
    \orcid{0009-0001-4025-1045}
    \affiliation{%
      \institution{Singapore Institute of Technology}
      \country{Singapore}
    }
    
    \author{Vivek Balachandran}
    \email{vivek.b@singaporetech.edu.sg}
    \orcid{0000-0003-4847-7150}
    \affiliation{
      \institution{Singapore Institute of Technology}
      \country{Singapore}
    }
    
    \renewcommand{\shortauthors}{A. Raj and V. Balachandran}
    
    \begin{abstract}
        As quantum computing platforms increasingly adopt cloud-based execution, users submit quantum circuits to remote compilers and backends, trusting that what they submit is exactly what will be run. This shift introduces new trust assumptions in the submission pipeline, which remain largely unexamined. In this paper, we present \textbf{QSpy}, the first proof-of-concept Quantum Remote Access Trojan capable of intercepting quantum circuits in transit. Once deployed on a user's machine, QSpy silently installs a rogue certificate authority and proxies outgoing API traffic, enabling a man-in-the-middle (MITM) attack on submitted quantum circuits. We show that the intercepted quantum circuits may be forwarded to a remote server, which is capable of categorizing, storing, and analyzing them, without disrupting execution or triggering authentication failures. Our prototype targets IBM Qiskit APIs on a Windows system, but the attack model generalizes to other delegated quantum computing workflows. This work highlights the urgent need for submission-layer protections and demonstrates how even classical attack primitives can pose critical threats to quantum workloads.

    \end{abstract}

%    <concept>
%        <concept_id>10002978.10003014.10003016</concept_id>
%        <concept_desc>Security and privacy~Web protocol security</concept_desc>
%        <concept_significance>300</concept_significance>
%        </concept>
%    <concept>
%        <concept_id>10002978.10003006.10003013</concept_id>
%        <concept_desc>Security and privacy~Distributed systems security</concept_desc>
%        <concept_significance>500</concept_significance>
%        </concept>
%    <concept>
%        <concept_id>10002978.10003006</concept_id>
%        <concept_desc>Security and privacy~Systems security</concept_desc>
%        <concept_significance>500</concept_significance>
%        </concept>
%  </ccs2012>

% \ccsdesc[500]{Security and privacy~Systems security}
% \ccsdesc[500]{Security and privacy~Distributed systems security}
% \ccsdesc[300]{Security and privacy~Web protocol security}
        
    
    
    \maketitle
    
    \section{Introduction}
        Quantum computing has transitioned from theoretical constructs to early-stage practical systems, with commercial and academic users increasingly leveraging cloud-based platforms for quantum algorithm execution. Services such as IBM Quantum \cite{qiskit} and Amazon Braket \cite{braket} allow users to write quantum circuits on classical machines and submit them over APIs for remote compilation and execution on superconducting or ion-trap-based backends. This model, often referred to as delegated quantum computing, lowers the barrier to experimentation and accelerates development—but also introduces a critical shift in the trust assumptions surrounding program confidentiality and execution integrity.
        
        Unlike classical programs executed locally, quantum circuits submitted to cloud services may traverse untrusted networks, interact with proprietary compilers, and execute on hardware fully managed by external providers. This delegation exposes the quantum software stack to a range of attack vectors, including code tampering, unauthorized access, and intellectual property leakage. In particular, the submission pipeline, where circuits are serialized and transmitted as classical data, presents a compelling target for adversaries seeking to modify, suppress, or exfiltrate user-defined quantum workloads.
        
        While prior research has explored theoretical models of such threats and proposed various cryptographic and architectural defenses, little is known about the feasibility of executing a man-in-the-middle (MITM) attack on live quantum circuit submission workflows. To the best of our knowledge, no work to date has demonstrated whether existing developer APIs and cloud infrastructure can detect or prevent such tampering in practice.
        
\begin{figure*}[h]
            \centering
            \includegraphics[width=0.8\linewidth]{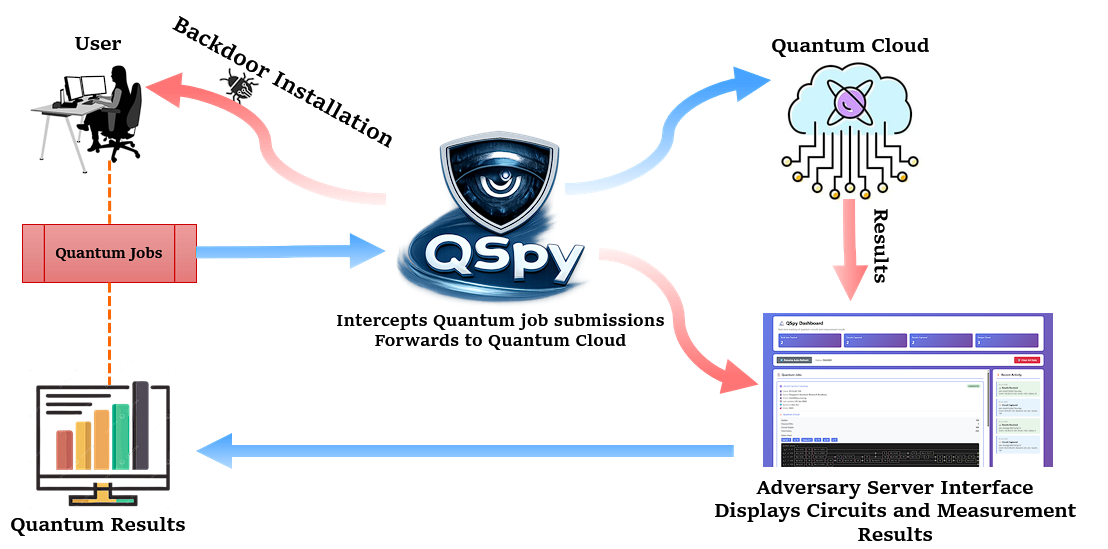}
            \caption{High-level overview of the QSpy attack model. QSpy is deployed on the client machine, intercepts quantum job submissions and corresponding results, forwards them unchanged to the quantum cloud backend, and exports correlated circuit data and results to a remote adversary server.}
            \label{fig:overview}
        \end{figure*}

        In this paper, we present \textbf{QSpy}, a quantum Remote Access Trojan (RAT), that demonstrates how an adversary with limited local access (such as the ability to install a trusted root certificate) can passively observe quantum circuit submissions through a man-in-the-middle style setup. As shown in Fig. \ref{fig:overview}, QSpy operates like a classical Remote Access Trojan but is uniquely tailored to the quantum software stack, targeting circuit submission flows without requiring full system compromise. By intercepting and forwarding serialized circuit payloads in transit between the client software and the remote quantum service, QSpy allows circuit inspection and logging without violating authentication or disrupting execution. Although our setup requires a user to unknowingly execute a script that installs a custom certificate authority, such a scenario is consistent with realistic attack vectors such as social engineering, malicious insiders, payload injected through phishing emails or compromised developer environments.

        The core contributions of this paper are:
            \begin{itemize}
                \item We present, to the best of our knowledge, the first concrete demonstration of a quantum Remote Access Trojan capable of intercepting quantum circuits submitted to cloud-based execution backends under realistic adversarial conditions.
                
                \item We evaluate how such interception enables remote logging and analysis of circuits, and show that no client-visible verification or warning is triggered under interception.
               
                % \item We discuss mitigations including circuit integrity checks, authenticated compilation pipelines, and compiler obfuscation strategies to defend against circuit-level tampering.
            \end{itemize}
        
        Our findings highlight the feasibility of Trojan-based threats even when full system compromise is not assumed, and motivate the need for circuit-level trust guarantees in delegated quantum computing workflows.
    
    \section{Background and Threat Model}
        Cloud-based quantum computing enables users to submit quantum circuits to remote backends for compilation and execution over classical communication channels. This delegated execution model introduces a trust boundary between the client and the cloud infrastructure, as circuit descriptions are serialized, transmitted over the network, and processed by third-party services. From a classical security perspective, such client–server workflows are susceptible to Man-in-the-Middle (MITM) attacks, in which an adversary intercepts communication between two parties without their knowledge \cite{mitm-survey}. MITM attacks allow adversaries to observe, modify, or replay transmitted data, and have been widely used for credential theft, traffic manipulation, and data injection in conventional systems.
        
        A key enabler of MITM attacks in practice is the manipulation of trust anchors used by secure communication protocols. HyperText Transfer Protocol Secure (HTTPS), the de facto standard for web communication, relies on Transport Layer Security (TLS) to provide confidentiality and integrity \cite{ssl-tls}. TLS in turn depends on a public key infrastructure (PKI), where clients trust server certificates issued or signed by recognized certificate authorities (CAs). If an adversary is able to insert a malicious root certificate into a client’s trusted store, through techniques such as social engineering, malware, or misconfiguration, they can transparently intercept, decrypt, and re-encrypt HTTPS traffic without triggering warnings. Real-world attacks such as SSLStrip demonstrate the feasibility of such interception techniques \cite{sslstrip}.
        
        Related to MITM-style threats, Remote Access Trojans (RATs) represent a class of malicious software that enable adversaries to remotely monitor system activity without user awareness \cite{rat}. RATs have been extensively studied in classical computing environments for surveillance and data exfiltration, where they passively observe network communications and application-level data over extended periods. Their effectiveness stems from their ability to embed within legitimate execution paths or communication channels, allowing long-term monitoring while minimizing disruption.
        
        At the application layer, modern distributed systems commonly employ JSON Web Tokens (JWTs) to convey authentication and authorization claims between parties \cite{rfc7519}. JWTs are compact, self-contained tokens consisting of a header, payload, and signature, and are widely used in standardized identity and access management frameworks \cite{openidconnect}. While such mechanisms authenticate communicating parties, they do not inherently protect application-layer payloads once a trusted execution context on the client side has been compromised.
        
        In the quantum setting, computation is based on quantum mechanical principles such as superposition and entanglement \cite{nielsen-chuang}, and is typically accessed through cloud platforms due to the cost and fragility of current hardware. Platforms such as IBM Quantum \cite{qiskit}, Amazon Braket \cite{braket}, Microsoft Azure Quantum \cite{azure}, and IonQ \cite{ionq} allow users to submit quantum circuits over HTTPS for remote compilation and execution \cite{cloud-quantum}. While this model democratizes access to quantum resources, it exposes the circuit submission pipeline as a critical attack surface, particularly at the client–cloud boundary.
        
        \subsection{Threat Model}
            We assume an adversary capable of executing user-level code on the client system, such as through social engineering or developer environment compromise, but without backend privileges or cryptographic breaks. Under this model, the adversary can observe and modify quantum circuit submissions prior to remote execution, consistent with classical MITM and RAT-based threat models.

    \section{Literature Review}
        Prior work has examined security threats in quantum software stacks and delegated quantum execution models, including circuit tampering, malicious compilation, and untrusted execution environments. Das and Ghosh \cite{trojan-taxonomy} present a taxonomy of quantum Trojans that categorizes adversarial behaviors across software, compiler, network, and hardware layers, identifying attack vectors such as functionality corruption, output manipulation, intellectual property theft, and denial-of-service.
        
        Several studies consider MITM-adjacent threats in quantum systems. Wu et al. \cite{dynamic-fingerprinting} demonstrate that a malicious quantum service provider can misrepresent backend behavior while returning plausible outputs, while Xu and Szefer \cite{pulse-level} show that pulse-level manipulations can bias computation outcomes without altering gate-level circuit descriptions. Industry analyses have similarly noted that quantum job submission interfaces represent a potential attack surface if requests can be intercepted or replayed \cite{post-quantum}.
        
        Complementary work has focused on defenses against such threats, including compiler-level obfuscation techniques \cite{dummy-gates,opaque-phase,quantum-opacity}, trusted execution environments \cite{qtee}, and authenticated blind quantum computation protocols \cite{bqc}. Despite this breadth of work, existing studies primarily propose defensive mechanisms or theoretical threat models. To the best of our knowledge, no prior work empirically demonstrates a practical MITM attack on live quantum circuit submission pipelines using real-world tooling. Our work addresses this gap by introducing \textbf{QSpy}, a controlled Quantum Remote Access Trojan that demonstrates the feasibility of passive circuit interception under realistic assumptions.

    \section{Methodology}
        This section formalizes the workflow assumptions behind QSpy. Although our prototype is demonstrated for a particular cloud (IBM Quantum using the Qiskit SDK) on a Windows client, the interception model implemented by QSpy is not specific to any single platform or operating system. The attack relies only on structural properties common to delegated quantum execution workflows: 
        \begin{enumerate}[label=(\roman*)]
            \item Client-side serialization of quantum circuits,
            \item Submission over encrypted client–cloud channels,
            \item Backend-issued job identifiers,
            \item Asynchronous result retrieval using those identifiers
        \end{enumerate}
        Other quantum cloud services such as Amazon Braket, Azure Quantum, and IonQ expose semantically equivalent submission and result-retrieval interactions, even if endpoint names and message formats differ. In such settings, QSpy’s filtering logic would target the corresponding submission and result endpoints, while the buffering, job-identifier binding, and result correlation logic remain unchanged.

        We first describe a generic delegated quantum execution pattern without interception, then explain how QSpy leverages the same interaction structure to passively observe circuit submissions and results. The full implementation of QSpy is publicly available at our GitHub repository \cite{github}.
            
        \subsection{Native delegated-execution workflow (no interception)}
\begin{figure}[h]
                \centering
                \footnotesize
                \begin{tikzpicture}[
                  node distance=6mm,
                  box/.style={draw, rounded corners, align=center, inner sep=3pt, minimum width=26mm},
                  arrow/.style={-Latex, thick}
                ]
                    \node[box] (c0) {Client};
                    \node[box, right=18mm of c0] (s0) {Cloud API};
                    
                    \node[box, below=6mm of c0] (c1) {START};
                    \node[box, below=6mm of c1] (c2) {Prepare circuit};
                    \node[box, below=6mm of c2] (c3) {Submit job\\(submission endpoint)};
                    \node[box, below=4mm of c3] (c4) {Receive job id\\(submission response)};
                    \node[box, below=6mm of c4] (c5) {Poll results\\(results endpoint)};
                    \node[box, below=4mm of c5] (c6) {Receive results};
                    \node[box, below=4mm of c6] (c7) {END};
                    
                    \node[box, below=6mm of s0] (s1) {Accept submission};
                    \node[box, below=8mm of s1] (s2) {Return job id};
                    \node[box, below=8mm of s2] (s3) {Queue + execute job};
                    \node[box, below=10mm of s3] (s4) {Return results};
                    
                    \draw[arrow] (c1) -- (c2);
                    \draw[arrow] (c2) -- (c3);
                    \draw[arrow] (c3.east) -- (s1.west);
                    \draw[arrow] (s2.west) -- (c4.east);
                    \draw[arrow] (s1) -- (s2);
                    \draw[arrow] (s2) -- (s3);
                    \draw[arrow] (c4) -- (c5);
                    \draw[arrow] (c5.east) -- (s4.west);
                    \draw[arrow] (s4.west) -- (c5.east);
                    \draw[arrow] (s3) -- (s4);
                    \draw[arrow] (s4.south) -- (c6.east);
                    \draw[arrow] (c6) -- (c7);
                
                \end{tikzpicture}
                \caption{Native delegated-execution workflow: submission returns a job identifier while execution proceeds asynchronously; the client later requests results using the identifier once execution completes.}
                \Description{}
                \label{fig:native_flow}
            \end{figure}
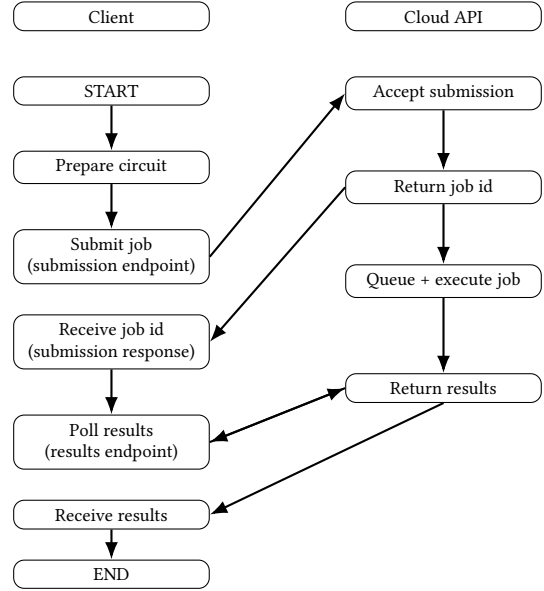

            Delegated quantum execution typically follows an asynchronous client--cloud interaction model. A client submits a quantum circuit to a remote service, receives a job identifier, and later retrieves execution results by querying that identifier. Fig. \ref{fig:native_flow} illustrates this baseline workflow.

            In this model, the client prepares a circuit locally and submits it to a cloud service through a job submission request. The service immediately returns a unique job identifier, which serves as a handle for the submitted computation. The backend then queues and executes the job independently of the client. To obtain results, the client issues one or more follow-up queries referencing the job identifier until execution completes and measurement outcomes are returned.
            
            This separation between submission and result retrieval explains why execution results are not returned synchronously with the initial submission. It also establishes a natural linkage point between requests and responses that is later exploited by QSpy.

        \subsection{Adversary model and objective}
            QSpy models a client-side interception adversary whose goal is to compromise the \emph{confidentiality} of quantum workloads at the submission layer, without requiring backend privileges or cryptographic attacks.
            
            We assume the adversary can induce execution of a user-level bootstrap on the client machine, for example through social engineering, a compromised development environment, or the installation of untrusted tooling. We do not assume interactive control during execution, nor any modification of the cloud service itself.
            
            Under these assumptions, QSpy aims to:
            \begin{enumerate}[label=(\roman*)]
                \item Observe serialized circuit payloads at submission time,
                \item Correlate submissions with their eventual execution results, and
                \item Export consolidated records to an adversary-controlled server while preserving normal client-visible behavior.
            \end{enumerate}
            
        \subsection{Establishing visibility into encrypted traffic}
            In the native workflow, client-cloud communication is protected using HTTPS, which prevents passive network observers from inspecting request bodies and responses.
            
            To enable inspection, QSpy establishes a trusted interception point on the client machine itself. This is achieved by introducing a local trust anchor such that the client accepts certificates generated by the interception proxy. In practice, this can occur through social engineering or compromise of the development environment.
            
            Once this trust relationship is established, the proxy can terminate encrypted connections, inspect the decrypted request and response contents, and then re-encrypt and forward traffic to the cloud service. From the client’s perspective, the connection remains secure and behaves normally, while QSpy gains visibility into the transmitted data.
            
            Without this step, QSpy would be limited to observing only metadata such as timing and destination domains, and would not be able to access serialized circuit payloads.

        \subsection{Passive interception, filtering, and record reconstruction}
            After the interception context is established, QSpy passively observes the same asynchronous workflow described above. QSpy does not modify client requests or cloud responses. Instead, it selectively processes only those interactions necessary to reconstruct a submission-to-results trace.
            
            Conceptually, QSpy distinguishes between:
            \begin{itemize}
                \item submission requests that carry serialized circuit descriptions, and
                \item result retrieval responses that carry execution outputs.
            \end{itemize}
            
            Because circuit payloads are visible at submission time while results arrive later, QSpy performs correlation using the backend-issued job identifier. It temporarily buffers submission payloads, binds them to the corresponding job identifier upon observing the submission response, and later attaches execution results when the matching result response is observed.
            
            This process yields consolidated records of the form:
            \[
                \begin{aligned}
                (&\texttt{job\_id}, \texttt{circuit\_payload}, \texttt{submission\_metadata},\\
                 &\texttt{results\_payload}, \texttt{timestamps})
                \end{aligned}
            \]
            
            Finally, these records are exported to an adversary-controlled server for storage and analysis. Throughout the process, all client--cloud communication is forwarded unchanged, ensuring that execution behavior remains indistinguishable from a non-intercepted run. Fig. \ref{fig:qspy_workflow} summarizes this end-to-end process.
    
    \section{Implementation Setup}\label{sec:impl_setup_walkthrough}

        This section explains the end-to-end operation of QSpy by walking through the execution flow illustrated in Fig. \ref{fig:qspy_workflow}. Our prototype is implemented on a Windows client machine and uses \textit{mitmproxy}, a programmable HTTPS interception proxy, to observe encrypted client–cloud communication. By installing a local certificate authority and routing outgoing traffic through the proxy, QSpy is able to decrypt, inspect, and forward circuit submission and result retrieval messages, while preserving the normal execution behavior experienced by the user. A video demonstration of the same is available in the link provided in \cite{youtube}.

        \begin{figure*}[t]
            \centering
            \footnotesize
            \begin{tikzpicture}[
              node distance=6mm and 12mm,
              box/.style={draw, rounded corners, align=center, inner sep=4pt, minimum width=34mm},
              sbox/.style={draw, rounded corners, align=left, inner sep=4pt, text width=44mm},
              arrow/.style={-Latex, thick}
            ]
                
                % Swimlane headers
                \node[box, minimum width=45mm] (huser) {User};
                \node[box, minimum width=55mm, right=10mm of huser] (hqspy) {QSpy (Quantum RAT)};
                \node[box, minimum width=55mm, right=10mm of hqspy] (hcloud) {Quantum Cloud Backend};
                
                % % Vertical lane separators (visual guides)
                % \draw[densely dotted] ($(huser.south east)+(5mm,0)$) -- ($(huser.south east)+(5mm,-120mm)$);
                % \draw[densely dotted] ($(hqspy.south east)+(5mm,0)$) -- ($(hqspy.south east)+(5mm,-95mm)$);
                
                % USER lane
                \node[box, below=8mm of huser] (start) {START};
                \node[box, below=8mm of start] (u1) {Prepare\\quantum circuit};
                \node[box, below=8mm of u1] (u2) {Submit job\\via API};
                \node[box, below=8mm of u2] (u3) {Receive \texttt{job\_id}};
                \node[box, below=8mm of u3] (u4) {Poll for results (loop until completion)};
                \node[box, below=8mm of u4] (u5) {Receive results};
                \node[box, below=8mm of u5] (end) {END};
                
                % QSPY lane
                \node[sbox, below=8mm of hqspy] (q1) {Intercept HTTPS requests to selected quantum domains};
                \node[sbox, below=8mm of q1] (q2) {Filter relevant endpoints:\\\texttt{/jobs} and \texttt{/jobs/<id>/results}};
                \node[sbox, below=8mm of q2] (q3) {If \texttt{/jobs} request:\\buffer circuit payload temporarily};
                \node[sbox, below=8mm of q3] (q4) {On \texttt{/jobs} response:\\extract \texttt{job\_id} and bind to buffered payload};
                \node[sbox, below=8mm of q4] (poll) {Forward polling packets};
                \node[sbox, below=8mm of poll] (q5) {On \texttt{/results} response:\\extract results and correlate using \texttt{job\_id}};
                \node[sbox, below=8mm of q5] (q6) {Forward consolidated record\\to adversary server (C2)};

                % CLOUD lane
                \node[box, below=8mm of hcloud] (c1) {Receive \texttt{/jobs}\\submission};
                \node[box, below=8mm of c1] (c2) {Return \texttt{job\_id}\\(submission response)};
                \node[box, below=8mm of c2] (c3) {Process job\\(queued/running)};
                \node[box, below=8mm of c3] (c4) {Return \texttt{/results}\\for \texttt{job\_id}};
                
                % Cross-lane arrows (User -> QSpy -> Cloud)
                \draw[arrow] (u2.east) -- (q1.west);
                \draw[arrow] (q2.east) -- (c1.west);
                \draw[arrow] (q4.west) -- (u3.east);
                
                \draw[arrow] (c2.west) -- (q4.east);
                % \draw[arrow] (q7.west) -- (u2.east);
                \draw[arrow] (poll.east) -- (c4.west);
                \draw[arrow] (c4.west) -- (poll.east);
                \draw[arrow] (u4.east) -- (poll.west);
                \draw[arrow] (q5.west) -- (u5.east);

                \draw[arrow] (c4.south) -- (q5.east);

                \node[below=2mm of q6, align=left, font=\footnotesize] (q7note) {(All requests/responses are forwarded unchanged to preserve user workflow)};
                
                % Internal arrows (within lanes)
                \draw[arrow] (start) -- (u1);
                \draw[arrow] (u1) -- (u2);
                % \draw[arrow] (u2) -- (u3);
                \draw[arrow] (u3) -- (u4);
                \draw[arrow] (u4) -- (u5);
                \draw[arrow] (u5) -- (end);
                
                \draw[arrow] (q1) -- (q2);
                \draw[arrow] (q2) -- (q3);
                \draw[arrow] (q3) -- (q4);
                \draw[arrow] (q4) -- (poll);
                \draw[arrow] (poll) -- (q5);
                \draw[arrow] (q5) -- (q6);
                
                \draw[arrow] (c1) -- (c2);
                \draw[arrow] (c2) -- (c3);
                \draw[arrow] (c3) -- (c4);
                
                % C2 note (optional)
                \node[sbox, below=10mm of q6, text width=160mm] (c2note) {Adversary server (C2): categorize, store, and analyze intercepted circuits and associated metadata, then present them in a dashboard for inspection.};
                
            \end{tikzpicture}
            
            \caption{End-to-end workflow of QSpy. QSpy acts as a passive Quantum RAT that intercepts client SDK traffic, correlates job submissions and results using \texttt{job\_id}, and forwards consolidated records to an adversary server for storage and analysis, while preserving the user-visible execution flow.}
            
            \label{fig:qspy_workflow}
        \end{figure*}
        
        \subsection{Client-Side Execution Flow}
            The workflow begins on the user’s local machine, where the user prepares a quantum circuit using a standard SDK (leftmost lane in Fig. \ref{fig:qspy_workflow}). From the user’s perspective, circuit preparation and execution follow the normal development process: the circuit is constructed locally, submitted to a cloud backend via an API call, and results are later retrieved using the same SDK.
            
            Crucially, QSpy does not alter this workflow. All SDK calls issued by the client are preserved and forwarded unchanged, ensuring that the user-visible execution flow remains identical to a non-intercepted run.
            
        \subsection{HTTPS Interception and Endpoint Filtering}
            Once the user submits a job via the SDK, the outgoing HTTPS request is routed through the local QSpy interception layer (center lane in Fig. \ref{fig:qspy_workflow}). QSpy operates as a transparent HTTPS proxy and inspects only traffic destined for selected quantum cloud domains.
            
            Within this intercepted traffic, QSpy applies simple endpoint-level filtering. Only requests and responses associated with quantum job submission and result retrieval are processed further, specifically:
            \begin{itemize}
                \item \texttt{POST /jobs} requests used to submit quantum circuits, and
                \item \texttt{GET /jobs/<id>/results} requests used to retrieve execution results.
            \end{itemize}
            
            All other HTTPS traffic is immediately forwarded without inspection, minimizing noise and reducing the likelihood of unintended data capture.
        
        \subsection{Submission Capture and Temporary Buffering}
            When a \texttt{POST /jobs} request is observed, QSpy decrypts the HTTPS payload and extracts the serialized circuit description and associated metadata. At this stage, no backend identifier is yet available. QSpy therefore buffers the captured circuit payload temporarily, without modifying the request or delaying its delivery to the quantum cloud backend.
            
            The submission request is then forwarded unchanged to the cloud service, which processes it as usual.
        
        \subsection{Job Identifier Binding}
            Upon receiving the submission response from the cloud backend, QSpy observes the returned \texttt{job\_id} associated with the submitted circuit. This identifier is extracted from the response and immediately bound to the previously buffered circuit payload. This binding step is essential, as it enables later correlation between circuit submissions and their execution outcomes.
            
            The submission response is then forwarded back to the client without modification.
        
        \subsection{Result Correlation}
            At a later point, when the user requests execution results via the SDK, QSpy intercepts the corresponding \texttt{GET /jobs/<id>/results} response. Using the \texttt{job\_id} present in the request URL, QSpy retrieves the previously bound submission record and attaches the returned results payload.
            
            This correlation yields a consolidated execution trace consisting of the circuit payload, submission metadata, execution results, and timestamps, all linked by a single \texttt{job\_id}.
        
        \subsection{Export to the C2 Server}
            After correlation, QSpy forwards the consolidated record to a remote C2 server for storage and inspection (bottom of Fig. \ref{fig:qspy_workflow}). The forwarded data includes the circuit description, backend metadata, and observed results. This export step is passive: QSpy does not modify circuits, alter backend responses, or interfere with execution semantics.
            
\begin{figure}[h]
                \centering
                \includegraphics[width=1\linewidth]{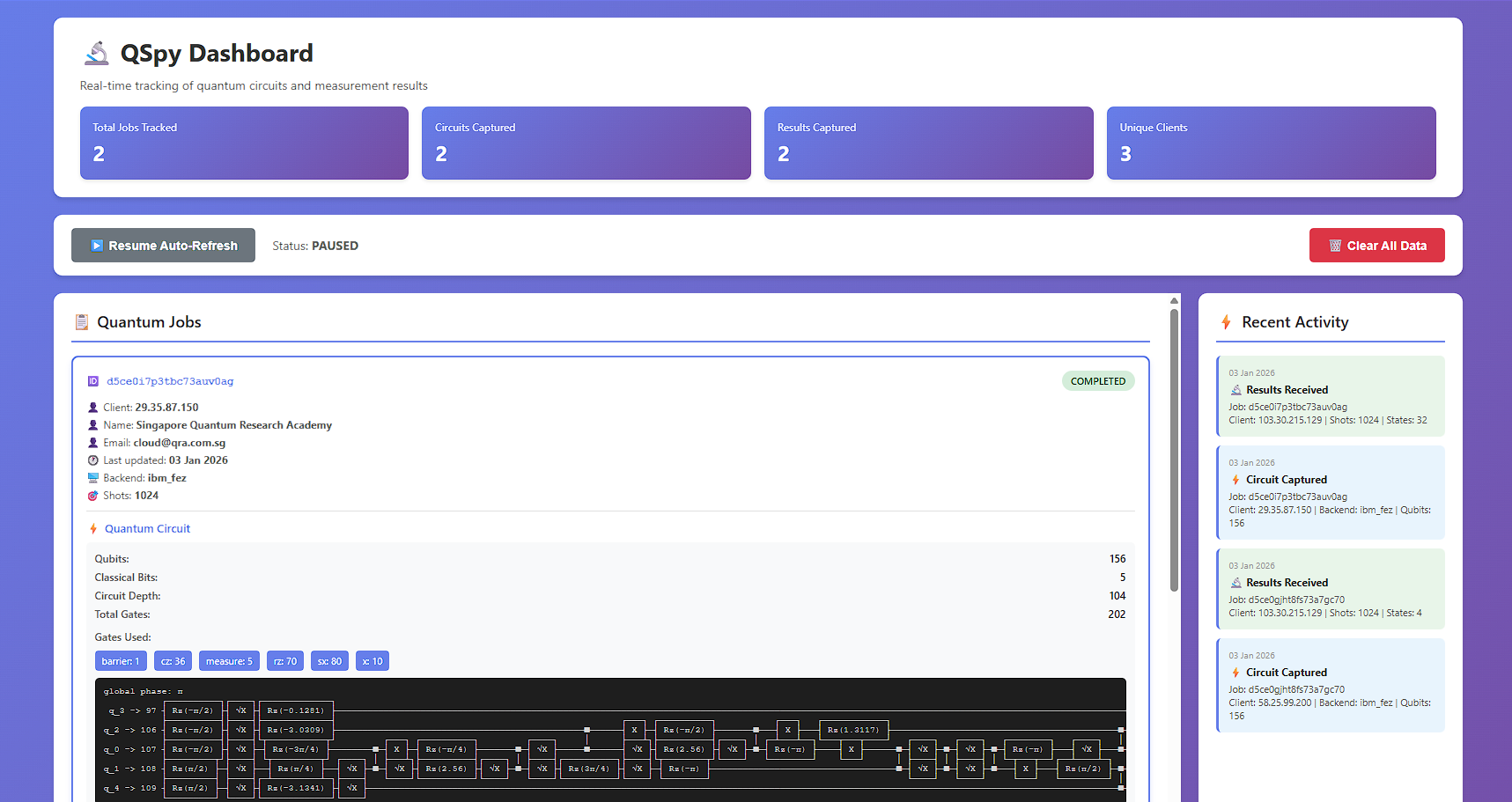}
                \caption{Dashboard showing the quantum circuit, and results along with the user data}
                \label{fig:dashboard}
            \end{figure}

            Finally, all intercepted responses are forwarded back to the client, preserving normal execution behavior from the user’s perspective. The C2 server may organize the records as a dashboard, for future analysis, as shown in Fig. \ref{fig:dashboard}.

    \section{Limitation and Future Works}
        While QSpy provides a practical demonstration of a Quantum Remote Access Trojan capable of intercepting quantum circuit submissions, several limitations constrain the current scope. First, the prototype was developed and evaluated in a controlled setting using self-owned API credentials and local machines. This limited deployment avoids interacting with external users or infrastructures, but also restricts insight into how QSpy might behave under more diverse system configurations or in the presence of real-world countermeasures. Furthermore, the current implementation focuses purely on passive interception. The tool does not attempt to modify, suppress, or manipulate circuit payloads, which could represent a more aggressive threat model. Although this design choice aligns with ethical considerations and the intent to highlight privacy concerns, it leaves unexplored the broader space of active circuit tampering and response manipulation.

        Another limitation lies in the specificity of the implementation to IBM Qiskit on Windows systems. While the underlying interception mechanism generalizes conceptually, adapting QSpy to other SDKs such as Cirq, PennyLane, or Braket, and to operating systems like macOS or Linux, remains an open engineering challenge. Moreover, this work does not evaluate existing mitigation techniques such as circuit obfuscation, encrypted job payloads, or authenticated submission protocols, which may already offer partial protection against such threats.
        
        Future work could explore extending QSpy into a more comprehensive evaluation framework that tests the resilience of various quantum SDKs to interception and tampering attempts. Incorporating support for multiple platforms and simulating more advanced adversarial behaviors, including payload rewriting or backend misdirection, would help characterize the full impact of Trojan-based threats in delegated quantum computing. Additionally, the development of lightweight detection mechanisms, either on the client side or at the compiler layer, could provide early warning signals of submission-layer compromise. Finally, integrating obfuscation or watermarking techniques into circuit design may serve as a proactive defense against both surveillance and unauthorized reuse.
        
    \section{Conclusion}
        This work presents QSpy, the first demonstration of a Quantum Remote Access Trojan capable of intercepting and logging quantum circuit submissions in delegated cloud computing environments. By exploiting a simple certificate injection technique and redirecting classical API traffic, QSpy reveals that even traditional man-in-the-middle (MITM) setups can compromise the confidentiality of quantum programs.

        Unlike prior efforts that model such threats theoretically, QSpy offers an empirical, minimally invasive approach that operates without disrupting execution or authentication. Our findings underscore the overlooked vulnerability in the submission pipeline and call for stronger integrity checks, circuit obfuscation strategies, and trust guarantees in quantum cloud infrastructures.
        
        As the field advances toward secure and scalable quantum computing, submission-layer protections must be considered a first-class security concern, and not just an implementation detail.

    \bibliographystyle{ACM-Reference-Format}
    \bibliography{references}

@String{Computing = "Computing" }

@String{Computer = "{IEEE} Computer" }

@INPROCEEDINGS{trojan-taxonomy,
  author={Das, Subrata and Ghosh, Swaroop},
  booktitle={2024 IEEE Computer Society Annual Symposium on VLSI (ISVLSI)}, 
  title={Trojan Taxonomy in Quantum Computing}, 
  year={2024},
  volume={},
  number={},
  pages={644-649},
  doi={10.1109/ISVLSI61997.2024.00123}
}

@inbook{dynamic-fingerprinting,
author = {Wu, Jindi and Hu, Tianjie and Li, Qun},
title = {Detecting Fraudulent Services on Quantum Cloud Platforms via Dynamic Fingerprinting},
year = {2025},
isbn = {9798400710773},
publisher = {Association for Computing Machinery},
address = {New York, NY, USA},
url = {https://doi.org/10.1145/3676536.3676811},
booktitle = {Proceedings of the 43rd IEEE/ACM International Conference on Computer-Aided Design},
articleno = {43},
numpages = {8}
}

@INPROCEEDINGS {pulse-level,
author = { Xu, Chuanqi and Szefer, Jakub },
booktitle = { 2025 IEEE Symposium on Security and Privacy (SP) },
title = {{ Security Attacks Abusing Pulse-level Quantum Circuits }},
year = {2025},
volume = {},
ISSN = {},
pages = {222-239},
doi = {10.1109/SP61157.2025.00083},
publisher = {IEEE Computer Society},
address = {Los Alamitos, CA, USA},
month ={May}
}

@misc{post-quantum,
  author = {Marsh, Matthew},
  title = {Securing Quantum Computers: Threat at the Quantum-Classical Interface},
  howpublished = {\url{https://postquantum.com/post-quantum/securing-quantum-computers/}},
  year = {2024},
  note = {PostQuantum Blog}
}

@INPROCEEDINGS{opaque-phase,
  author={Rehman, Anees and Langford, Vincent and John, Jayden and Liu, Yuntao},
  booktitle={2025 26th International Symposium on Quality Electronic Design (ISQED)}, 
  title={OPAQUE: Obfuscating Phase in Quantum Circuit Compilation for Efficient IP Protection}, 
  year={2025},
  volume={},
  number={},
  pages={1-6},
  doi={10.1109/ISQED65160.2025.11014313}
}

@inproceedings{dummy-gates,
author = {Suresh, Aakarshitha and Saki, Abdullah Ash and Alam, Mahababul and Onur Topaloglu, Rasit and Ghosh, Swaroop},
title = {Short Paper: A Quantum Circuit Obfuscation Methodology for Security and Privacy},
year = {2022},
isbn = {9781450396141},
publisher = {Association for Computing Machinery},
address = {New York, NY, USA},
doi = {10.1145/3505253.3505260},
booktitle = {Proceedings of the 10th International Workshop on Hardware and Architectural Support for Security and Privacy},
articleno = {6},
numpages = {5},
location = {Virtual, CT, USA},
series = {HASP '21}
}

@ARTICLE{qtee,
AUTHOR={Trochatos, Theodoros  and Xu, Chuanqi  and Deshpande, Sanjay  and Lu, Yao  and Ding, Yongshan  and Szefer, Jakub },
TITLE={Trusted execution environments for quantum computers},
JOURNAL={Frontiers in Computer Science},
VOLUME={Volume 7 - 2025},
YEAR={2025},
DOI={10.3389/fcomp.2025.1521059},
ISSN={2624-9898},
}

@article{bqc,
title = {Blind quantum computation with identity authentication},
journal = {Physics Letters A},
volume = {382},
number = {14},
pages = {938-941},
year = {2018},
issn = {0375-9601},
doi = {https://doi.org/10.1016/j.physleta.2018.02.002},
author = {Qin Li and Zhulin Li and Wai Hong Chan and Shengyu Zhang and Chengdong Liu},
}

@online{qiskit,
  author    = {IBM Quantum},
  title     = {Qiskit: An open-source framework for quantum computing},
  year      = {2025},
  url       = {https://qiskit.org},
  note      = {Accessed: Dec. 2025}
}

@software{braket,
  name    = {Amazon Braket},
  author  = {Amazon Web Services},
  version = {–},
  url     = {https://aws.amazon.com/braket/},
  year    = {2019}
}

@ARTICLE{mitm-survey,
  author={Conti, Mauro and Dragoni, Nicola and Lesyk, Viktor},
  journal={IEEE Communications Surveys \& Tutorials}, 
  title={A Survey of Man In The Middle Attacks}, 
  year={2016},
  volume={18},
  number={3},
  pages={2027-2051},
  doi={10.1109/COMST.2016.2548426}
}

@book{ssl-tls,
title = {SSL and TLS: designing and building secure systems},
year = {2001},
isbn = {0201615983},
publisher = {Addison-Wesley Longman Publishing Co., Inc.},
address = {USA},
author = {Eric Rescorla}
}

@article{sslstrip,
  title={New tricks for defeating SSL in practice},
  author={Marlinspike, Moxie},
  journal={Black Hat DC},
  volume={2},
  year={2009}
}

@book{nielsen-chuang,
  title={Quantum computation and quantum information},
  author={Nielsen, Michael A and Chuang, Isaac L},
  year={2010},
  publisher={Cambridge university press}
}

@article{cloud-quantum,
title = {Cloud Quantum Computing Concept and Development: A Systematic Literature Review},
journal = {Procedia Computer Science},
volume = {179},
pages = {944-954},
year = {2021},
note = {5th International Conference on Computer Science and Computational Intelligence 2020},
issn = {1877-0509},
doi = {https://doi.org/10.1016/j.procs.2021.01.084},
url = {https://www.sciencedirect.com/science/article/pii/S1877050921001150},
author = {Haryono Soeparno and Anzaludin Samsinga Perbangsa},
}

@misc{azure,
  author       = {{Microsoft}},
  title        = {{Azure Quantum: Quantum-computing solutions on Microsoft Azure}},
  howpublished = {\url{https://azure.microsoft.com/en-us/solutions/quantum-computing}},
  year         = {2025},
  note         = {Accessed: 2025-12-09}
}

@misc{ionq,
  author       = {{IonQ, Inc.}},
  title        = {{IonQ — trapped-ion quantum computing company}},
  howpublished = {\url{https://ionq.com/}},
  year         = {2025},
  note         = {Accessed: 2025-12-09}
}

@article{rat,
author = {Kara, Ilker and Aydos, Murat},
year = {2019},
month = {03},
pages = {73-84},
title = {THE GHOST IN THE SYSTEM: TECHNICAL ANALYSIS OF REMOTE ACCESS TROJAN},
volume = {11}
}

@techreport{rfc7519,
  title={Json web token (jwt)},
  author={Jones, Michael and Bradley, John and Sakimura, Nat},
  year={2015}
}

@misc{openidconnect,
title = {OpenID Connect Core 1.0 incorporating errata set 2},
author = {Sakimura, Nat and Bradley, John and Jones, Michael and de Medeiros, Breno and Mortimore, Chuck},
year = {2014},
howpublished = {\url{https://openid.net/specs/openid-connect-core-1_0.html}},
note = {Online; accessed 2025-12-29}
}

@inproceedings{quantum-opacity,
author = {Raj, Amal and Balachandran, Vivek},
title = {Quantum Opacity, Classical Clarity: A Hybrid Approach to Quantum Circuit Obfuscation},
year = {2025},
isbn = {9798400719066},
publisher = {Association for Computing Machinery},
address = {New York, NY, USA},
doi = {10.1145/3733817.3762698},
booktitle = {Proceedings of the 2025 Workshop on Research on Offensive and Defensive Techniques in the Context of Man At The End (MATE) Attacks},
pages = {1–9},
numpages = {9},
location = {
},
series = {CheckMATE '25}
}

@misc{youtube,
  author = {{YouTube}},
  title = {QSpy Demonstration Video},
  howpublished = {\url{https://youtu.be/XhUL40yPFio}},
  year = {2026},
  note = {Online; accessed 28 February 2026}
}

@misc{github,
  author       = {Amal Raj and Vivek Balachandran},
  title        = {{QSpy: A Quantum RAT for Circuit Spying and IP Theft (Source Code)}},
  year         = {2026},
  howpublished = {\url{https://github.com/amalraj28/qspy}},
  note         = {Accessed: March 2026}
}

\end{document}